\documentclass[12pt]{article}

\usepackage{amsmath}
\usepackage{amssymb}
\usepackage{epsfig}
\usepackage{graphicx}
\usepackage{color}

\def\e{\epsilon}

\def\e3{$\epsilon_3$}

\def\ch2{$\chi^2$}

\def\co#1{{\ifmmode{\cal O}_{#1}\else${\cal O}_{#1}$\fi}}

\newdimen\unit
\def\point#1 #2 #3{\vbox to0pt{\kern-#2\unit
  \hbox{\kern#1\unit#3}\vss}
 \nointerlineskip}

\newcommand{\be}{\begin{equation}}
\newcommand{\ee}{\end{equation}}
\newcommand{\bea}{\begin{eqnarray}}
\newcommand{\eea}{\end{eqnarray}}

\begin{document}
\thispagestyle{empty} \noindent
\begin{flushright}
        OHSTPY-HEP-T-05-002     \\
        UCD-05-08 \\
        July 2005
\end{flushright}

\vspace{1cm}
\begin{center}
  \begin{large}
  \begin{bf}

Bi-large Neutrino Mixing and CP violation in an SO(10) SUSY GUT for Fermion Masses

\end{bf}
 \end{large}
\end{center}
  \vspace{1cm}
     \begin{center}
R. Derm\' \i \v sek$^\dagger$ and S. Raby$^*$\\
      \vspace{0.3cm}
\begin{it}
$^\dagger$Department of Physics, University of California at
Davis, Davis, CA 95616 \\
$^*$Department of Physics, The Ohio State University, 191 W. Woodruff Ave., Columbus, Ohio  43210
\end{it}
  \end{center}
  \vspace{1cm}
\centerline{\bf Abstract}
\begin{quotation}
\noindent We construct a simple $SO(10)$ SUSY GUT with $D_3$ family symmetry and low energy R parity. The model
describes fermion mass matrices with 14 parameters and gives excellent fits to 20 observable masses and mixing
angles in both quark and lepton sectors, giving 6 predictions. Bi-large neutrino mixing is obtained with
hierarchical quark and lepton Yukawa matrices; thus avoiding the possibility of large lepton flavor violation.
The model naturally predicts small 1-3 neutrino mixing, $\sin \theta_{13} \simeq 0.05$, and a CP violating phase
$\delta$ close to $\pi/2$. Among other interesting predictions is a tiny effective Majorana mass for neutrinoless
double-beta decay. Leptogenesis is also possible with the decay of the lightest right-handed neutrino giving an
acceptable CP violating asymmetry $\epsilon_1$ of order $10^{-6}$, {\em and with the correct sign for the
resultant baryon asymmetry}.  Note, similar models with the non-abelian symmetry groups $SU(2)$ or $D_4$, instead
of $D_3$, can be constructed.

\end{quotation}

There are very few experimental indications for new physics beyond the extremely successful standard model.   By
far the two most exciting harbingers of new physics are the successful prediction of gauge coupling unification
in supersymmetric GUTs~\cite{sguts} and the discovery of neutrino masses and mixing.   Both of these experimental
results hint at a new large mass scale of order $10^{16}$ GeV.   In addition, SUSY GUTs can alleviate the gauge
hierarchy problem implying the exciting possibility for the discovery of supersymmetric partners at the LHC.   It
is thus important to see to what extent SUSY GUTs and neutrino masses and mixing angles are compatible.

The first problem one encounters in any theory with quark-lepton unification is the fact that flavor mixing in
the quark sector is small, while for neutrinos there are two large mixing angles. In the quark sector it is
well-known that hierarchical Yukawa matrices ``naturally" fit the family hierarchy, as well as the hierarchy of
flavor mixing evident in the CKM matrix. Moreover this can be described group theoretically in terms of the
hierarchical breaking of family symmetries via the Froggatt-Nielsen mechanism~\cite{Froggatt:1978nt}. In
supersymmetric theories, family symmetries play a dual role.  In addition to describing the hierarchy of fermion
masses and mixing, they ``naturally" align squark mass matrices with their quark superpartners;\footnote{For
non-abelian family symmetries, squark and slepton mass matrices are proportional to the unit matrix prior to
family symmetry breaking.  Then when the family symmetry is broken, generating the hierarchy of fermion masses,
the squark and slepton mass matrices also receive corrections which are aligned with their fermionic partners.}
thus ameliorating problems with flavor violation. In this letter, we show that bi-large neutrino mixing is
naturally incorporated into SUSY GUTs with hierarchical quark and lepton Yukawa matrices; thus avoiding the
possibility of large lepton flavor violation~\footnote{In several recent SUSY GUT constructions, bi-large
neutrino mixing is obtained with so-called {\em lopsided} Yukawa matrices for quarks and leptons \cite{lopsided}.
In this approach, use is made of the $SU(5)$ relation $Y_e = Y_d^T$ where large, {\em unobservable} right-handed
mixing for down quarks is converted to large left-handed mixing angles for leptons (for a review, see
\cite{King:2003jb,Altarelli:2004za}). In this case, one takes $(Y_d)_{23} \approx (Y_d)_{33} \sim 1$. One
apparent difficulty of this approach is the large induced lepton flavor violation for processes such as $\mu
\rightarrow e \gamma$ due to large off-diagonal $(Y_e)_{32} \approx (Y_e)_{33} \sim 1$ \cite{Barr:2003fn}.}.
Given Yukawa matrices of the form
\bea  Y \simeq \left( \begin{array}{ccc}  0 & b & b \\
b & a & a \\  b & a & 1 \end{array} \right) \lambda  \label{eq:ynu} \eea with $b \ll a \ll 1$ and a hierarchical
right-handed neutrino mass matrix of the form
\bea  M_R = \left( \begin{array}{ccc}  M_{R_1} & 0 & 0 \\
0 & M_{R_2} & 0 \\  0 & 0 & M_{R_3} \end{array} \right) \eea with  $M_{R_1} \ll M_{R_2} \ll M_{R_3}$ one
``naturally" obtains large neutrino
mixing~\cite{Smirnov:1993af,Altarelli:1999dg,Frampton:2002qc,Kuchimanchi:2002yu,King:2002nf,Dorsner:2004qb,Dermisek:2004tx}.
Neglecting, for example, the contribution of $M_{R_{2,3}}$ we find, upon integrating out the lightest
right-handed neutrino \bea {\cal M}_\nu = \left(
\begin{array}{ccc} 0 & 0 & 0 \\ 0 & 1 & 1 \\ 0 & 1 & 1
\end{array} \right) \ \frac{ (\lambda \ b \ v_u )^2 }{M_{R_1}} . \label{eq:maximal} \eea
giving maximal mixing angle for atmospheric neutrinos.   It has been shown by one of us that this mechanism also
gives acceptable bi-large neutrino mixing if the contributions of the lightest two neutrinos are
comparable~\cite{Dermisek:2004tx}.

Our model is an SO(10) realization of this general mechanism.   This letter is organized as follows: first we
present the superpotential for the fermion mass sector of an $SO(10)$ SUSY GUT.   We then fit the low energy data
using a global $\chi^2$ analysis.  The $\chi^2$ function includes 20 low energy observables for fermion masses
and mixing angles, in addition to the three low energy gauge couplings. We obtain an excellent fit to the data.
Then, we present predictions for additional observables (not included in the $\chi^2$ analysis): $\sin^2
\theta_{13}$, the CP violating angle $\delta$ (and Jarlskog parameter $J$), the effective neutrinoless double
beta decay mass parameter $\langle m \rangle_{\beta \beta}$, the tritium beta decay mass parameter
$m_{\nu_e}^{eff}$, and finally the lepton number asymmetry parameter relevant for leptogenesis, $\epsilon_1$.
Alternate versions of the model with family symmetries $SU(2)$ or $D_4$ (instead of $D_3$) can easily be
constructed~\cite{Dermisek:2005ij}.

\medskip
\noindent {\it The model }
\medskip

Consider an $SO(10)$ SUSY GUT with a $D_3 \times [U(1) \times Z_2 \times Z_3]$ family symmetry.\footnote{For
related charged fermion analyses in $SO(10)$ SUSY GUTS with $D_3 \times U(1)$ (or $SU(2) \times U(1)^n$) family
symmetries, see \cite{Dermisek:1999vy} (or \cite{Barbieri:1996ww,Blazek:1999ue,Raby:2003ay}).  Note, the novel
feature in this letter is the structure and results for neutrino masses and mixing.} The three families of quarks
and leptons are contained in three 16 dimensional representations of $SO(10)$ $ \{ 16_a, \; 16_3 \}$ with $16_a,
\; a = 1,2$ a $D_3$ flavor doublet (see appendices of Ref.~\cite{Dermisek:1999vy} for details on $D_3$). Consider
the charged fermion sector first. Although the charged fermion sector is not the main focus of this letter it is,
however, necessary to first construct a superpotential which is consistent with charged fermion masses and mixing
angles, since in $SO(10)$ the neutrino and charged fermion sectors are inextricably intertwined.

The superpotential resulting in charged fermion masses and mixing angles is given by
\begin{eqnarray} W_{ch. fermions} = & 16_3 \ 10 \ 16_3 +  16_a \ 10 \ \chi_a & \label{eq:chfermionD3}
\\ & +  \bar \chi_a \ ( M_{\chi} \ \chi_a + \ 45 \ \frac{\phi_a}{\hat M} \ 16_3 \ + \ 45 \ \frac{\tilde
\phi_a}{\hat M} \  16_a + {\bf A} \ 16_a ) & \nonumber
\end{eqnarray}
where $M_{\chi} = M_0 ( 1 + \alpha X + \beta Y )$ includes $SO(10)$ breaking vevs in the $X$ and $Y$ directions,
$\phi^a, \ \tilde \phi^a \, (D_3 \; {\rm doublets}),  \ {\bf A} \;  ({\bf 1_B} \; {\rm singlet })$ are $SO(10)$
singlet flavon fields, and $\hat M, \ M_0$ are $SO(10)$ singlet masses.  The fields $45, \; A, \ \phi, \ \tilde
\phi \;$ are assumed to obtain vevs $\langle 45 \rangle \sim (B - L) \ M_G$, $\; A \ll M_0 \;$ and
\begin{equation} \langle \phi \rangle =  \left( \begin{array}{c} \phi_1 \\
\phi_2 \end{array} \right), \; \langle \tilde \phi \rangle = \left( \begin{array}{c} 0 \\ \tilde \phi_2
\end{array} \right) \end{equation} with $\phi_1 > \phi_2$.

The superpotential, (Eqn. \ref{eq:chfermionD3}) results in the following charged fermion Yukawa
matrices:\footnote{In our notation, Yukawa matrices couple electroweak doublets on the left to singlets on the
right. It has been shown in Ref. \cite{Blazek:1999ue} that excellent fits to charged fermion masses and mixing
angles are obtained with this Yukawa structure.}
\begin{eqnarray}
Y_u =&  \left(\begin{array}{ccc}  0  & \epsilon' \ \rho & - \epsilon \ \xi  \\
             - \epsilon' \ \rho &  \tilde \epsilon \ \rho & - \epsilon     \\
       \epsilon \ \xi   & \epsilon & 1 \end{array} \right) \; \lambda & \nonumber \\
Y_d =&  \left(\begin{array}{ccc}  0 & \epsilon'  & - \epsilon \ \xi \ \sigma \\
- \epsilon'   &  \tilde \epsilon  & - \epsilon \ \sigma \\
\epsilon \ \xi  & \epsilon & 1 \end{array} \right) \; \lambda & \label{eq:yukawaD3} \\
Y_e =&  \left(\begin{array}{ccc}  0  & - \epsilon'  & 3 \ \epsilon \ \xi \\
          \epsilon'  &  3 \ \tilde \epsilon  & 3 \ \epsilon  \\
 - 3 \ \epsilon \ \xi \ \sigma  & - 3 \ \epsilon \ \sigma & 1 \end{array} \right) \; \lambda &
 \nonumber \end{eqnarray}
with  \begin{eqnarray}  \xi \;\; =  \;\; \phi_2/\phi_1; & \;\;
\tilde \epsilon  \;\; \propto   \;\; \tilde \phi_2/\hat M;  & \label{eq:omegaD3} \\
\epsilon \;\; \propto  \;\; \phi_1/\hat M; &  \;\;
\epsilon^\prime \;\; \sim  \;\;  ({\bf A}/M_0); \nonumber \\
  \sigma \;\; =   \;\; \frac{1+\alpha}{1-3\alpha}; &  \;\; \rho \;\; \sim   \;\;
  \beta \ll \alpha .& \nonumber \end{eqnarray}

Let us now discuss neutrino masses.  In the three $16$s we have three electroweak doublet neutrinos ($\nu_a, \
\nu_3$) and three electroweak singlet anti-neutrinos ($\bar \nu_a, \ \bar \nu_3$).\footnote{In an equivalent
notation, we have three left-handed neutrinos ($\nu_{L a} \equiv \nu_a, \; \nu_{L 3} \equiv \nu_3$) and three
right-handed neutrinos defined by ($\nu_{R a} \equiv \bar \nu_a^*, \ \nu_{R 3} \equiv \bar \nu_3^*$).}  The
superpotential $W_{ch. fermions}$ also results in a neutrino Yukawa matrix:
\begin{eqnarray}
Y_{\nu} =&  \left(\begin{array}{ccc}  0  & - \epsilon' \ \omega & {3 \over 2} \ \epsilon \ \xi \ \omega \\
      \epsilon'  \ \omega &  3 \ \tilde \epsilon \  \omega & {3 \over 2} \ \epsilon \ \omega \\
       - 3 \ \epsilon \ \xi \ \sigma   & - 3 \ \epsilon \ \sigma & 1 \end{array} \right) \; \lambda &
 \end{eqnarray} with $\omega \;\; =  \;\; 2 \, \sigma/( 2 \,
\sigma - 1)$ and a Dirac neutrino mass matrix given by
 \begin{equation} m_\nu \equiv Y_\nu \frac{v}{\sqrt{2}} \sin\beta.
 \label{eq:mnuD3}
  \end{equation}
In addition, the anti-neutrinos get GUT scale masses by mixing with three $SO(10)$ singlets $\{ N_a, \ a = 1,2;
\;\; N_3 \}$ transforming as a $D_3$ doublet and singlet respectively. The full superpotential is given by $W =
W_{ch. fermions} + W_{neutrino}$ with
\begin{eqnarray} \label{eq:WneutrinoD3} W_{neutrino} = & \overline{16} \left(\lambda_2 \ N_a \ 16_a \ + \ \lambda_3 \ N_3 \ 16_3 \right) & \\
& +  \;\; \frac{1}{2} \left(S_{a} \ N_a \ N_a \;\; + \;\; S_3 \ N_3 \ N_3\right).   & \nonumber
\end{eqnarray}
We assume $\overline{16}$ obtains a vev, $v_{16}$, in the right-handed neutrino direction, and $\langle S_{a}
\rangle = M_a$ for $a = 1,2$ (with $M_2 > M_1$) and $\langle S_3 \rangle = M_3$.\footnote{These are the most
general set of vevs for $\phi_a$ and  $S_{a}$.  The zero vev for $\tilde \phi_1$ can be enforced with a simple
superpotential term such as $S \ \tilde \phi_a \ \tilde \phi_a$.  }

We thus obtain the effective neutrino mass terms given by
\begin{equation} W =  \nu \ m_\nu \ \bar \nu + \bar \nu \ V \ N +
\frac{1}{2} \ N \ M_N \ N \end{equation} with
\begin{equation} V = v_{16} \ \left(
\begin{array}{ccc} 0 &  \lambda_2 & 0 \\
\lambda_2 & 0 & 0 \\ 0 & 0 &  \lambda_3 \end{array} \right), \; M_N = diag( M_1,\ M_2,\ M_3) . \end{equation}

A simple family symmetry giving the desired form of the superpotential\footnote{Note, highly suppressed terms of
the form $S_a \ N_a \ N_3 \ \tilde \phi_a \ \phi_a^2$ are still allowed.  This too can be forbidden by an
additional discrete $Z_4$ symmetry.} is $D_3 \times U(1) \times Z_2 \times Z_3$ where the $D_3$ charges were
defined earlier, while the $U(1)$ charge assignments are (1 for $16_3$, 2 for $16_a$, -2 for $N_a$, -1 for $N_3$,
-1 for 45, 0 for $\overline{16}$ and $\overline{\chi}_a$) and everyone else fixed by these. In addition we assign
$Z_2$ charges $(16_3, 16_a, N_3, N_a, \overline{\chi}_a, \chi_a )$ odd, all others even and $Z_3$ charges $\alpha
= e^{\frac{2\pi i}{3}}$ for all fields, except $45$ with charge 1. {\em Note, that $Z_2$ can also be interpreted
as a family reflection symmetry which guarantees an unbroken low energy R parity} \cite{Dimopoulos:1981dw}.

The electroweak singlet neutrinos $ \{ \bar \nu, N \} $ have large masses $V , M_N \sim M_G$.  After integrating
out these heavy neutrinos, we obtain the light neutrino mass matrix given by
\begin{equation}
{\cal M} =   m_\nu  \ M_R^{-1} \ m_\nu^T ,
\end{equation}
where the effective right-handed neutrino Majorana mass matrix is given by:
\begin{equation}
M_R =  V \ M_N^{-1}  \ V^T  \  \equiv \  {\rm diag} ( M_{R_1}, M_{R_2}, M_{R_3} ),
\end{equation}
with \bea M_{R_1} = (\lambda_2 \ v_{16})^2/M_2, \quad  M_{R_2} = (\lambda_2 \ v_{16})^2/M_1, \quad  M_{R_3} =
(\lambda_3 \ v_{16})^2/M_3 . \label{eq:rhmass} \eea  Defining $U_e$ as the $3\times3$ unitary matrix for
left-handed leptons needed to diagonalize $Y_e$ (Eqn. \ref{eq:yukawaD3}), i.e. $Y_e^D = U_e^T \ Y_e \ U_{\bar
e}^*$ and also $U_\nu$ such that $U_\nu^T \ {\cal M} \ U_\nu = {\cal M}_D = {\rm diag}( m_{\nu_1}, \ m_{\nu_2}, \
m_{\nu_3})$, then the neutrino mixing matrix is given by $U_{PMNS} = U_e^\dagger \ U_\nu$ in terms of the flavor
eigenstate ($\nu_\alpha$, $\alpha =  e, \ \mu, \ \tau$) and mass eigenstate ($\nu_i$, $i = 1,2,3$) basis fields
with \bea \nu_\alpha = \sum_i (U_{PMNS})_{\alpha i} \ \nu_i . \eea

For $U_{PMNS}$ we use the notation of Ref~\cite{Eidelman:2004wy} with  \bea \left( \begin{array}{c} \nu_e \\
\nu_\mu \\ \nu_\tau \end{array} \right) = \left(
\begin{array}{ccc} c_{1 2} c_{1 3} & s_{1 2} c_{1 3} & s_{1 3} e^{-i\delta} \\ -s_{12} c_{23} - c_{12} s_{23}
s_{13} e^{i\delta} & c_{12} c_{23} - s_{12} s_{23} s_{13} e^{i\delta} & s_{23} c_{13} \\ s_{12} s_{23} - c_{12}
c_{23} s_{13} e^{i\delta} & -c_{12} s_{23} - s_{12} c_{23} s_{13} e^{i\delta} & c_{23} c_{13} \end{array} \right)
\left(
\begin{array}{c} e^{i\alpha_1 /2} \nu_1 \\ e^{i\alpha_2 /2}  \nu_2 \\ \nu_3 \end{array} \right) \label{eq:PMNS}
\eea

The $3 \times 3$ Majorana mass matrix is of the general form discussed by many authors \cite{Dermisek:2004tx}. We
have
\begin{equation} {\cal M} =  {\cal P}_1 \ {\cal M}_1 \ {\cal P}_1+ {\cal P}_2 \ {\cal M}_2 \ {\cal P}_2  +
 {\cal P}_3 \ {\cal M}_3 \ {\cal P}_3
\end{equation} with
\begin{equation} {\cal M}_1 = \left( \begin{array}{ccc} 0 & 0 & 0 \\ 0 & (\epsilon' \ |\omega|)^2 & - 3 \
\epsilon' \ \epsilon \ |\xi \ \sigma \ \omega| \\ 0 & - 3 \ \epsilon' \ \epsilon \ |\xi \ \sigma \ \omega| & ( 3
\ \epsilon \ |\xi \ \sigma|)^2 \end{array} \right) \ \left(\frac{\lambda \ v \ \sin\beta}{\sqrt{2} \ \lambda_2  \
v_{16}}\right)^2 \ M_2 ;
\end{equation}
\begin{equation} {\cal M}_2 = \left( \begin{array}{ccc} (\epsilon' \ |\omega|)^2 & - 3 \ \epsilon' \
|\tilde \epsilon \ \omega^2| &
 3 \ \epsilon' \ \epsilon \ |\sigma \ \omega| \\ - 3 \ \epsilon' \ |\tilde \epsilon \ \omega^2| & (3 \ |\tilde \epsilon \ \omega|)^2 & -
9 \  \epsilon \ |\tilde \epsilon \ \sigma \ \omega| \\  3 \ \epsilon' \ \epsilon \ |\sigma \ \omega| & - 9 \
 \epsilon \ |\tilde \epsilon \ \sigma \ \omega| & ( 3 \ \epsilon \ |\sigma|)^2 \end{array} \right) \
\left(\frac{\lambda \ v \ \sin\beta}{\sqrt{2} \ \lambda_2 \ v_{16}}\right)^2 \ M_1 ;
\end{equation}
\begin{equation} {\cal M}_3 = \left( \begin{array}{ccc} (\frac{3}{2} \ \epsilon \ |\xi \ \omega|)^2 & |\xi| \
(\frac{3}{2} \ \epsilon \ |\omega|)^2 & \frac{3}{2} \ \epsilon \ |\xi \ \omega| \\ |\xi| \ (\frac{3}{2} \
\epsilon \ |\omega|)^2 & (\frac{3}{2} \ \epsilon \ |\omega|)^2 & \frac{3}{2} \ \epsilon \ |\omega| \\
\frac{3}{2} \ \epsilon \ |\xi \ \omega| & \frac{3}{2} \ \epsilon \ |\omega| & 1
\end{array} \right) \ \left(\frac{\lambda \ v \ \sin\beta}{\sqrt{2} \ \lambda_3 \ v_{16}}\right)^2 \ M_3
\end{equation}
and
\begin{equation} {\cal P}_1 = \left( \begin{array}{ccc} 1 & 0 & 0 \\ 0 & e^{-i \Phi_\omega} & 0 \\
 0 & 0 & e^{-i (\Phi_\sigma + \Phi_\xi)} \end{array} \right) ;
\end{equation}
\begin{equation} {\cal P}_2 = \left( \begin{array}{ccc} e^{-i \Phi_\omega} & 0 & 0 \\ 0 & e^{-i (\Phi_{\tilde \epsilon} +
\Phi_\omega)} & 0 \\
 0 & 0 & e^{-i \Phi_\sigma} \end{array} \right) ;
\end{equation}
\begin{equation} {\cal P}_3 = \left( \begin{array}{ccc} e^{-i (\Phi_\omega + \Phi_\xi)} & 0 & 0 \\ 0 & e^{-i \Phi_\omega} & 0 \\
 0 & 0 & 1 \end{array} \right).
\end{equation}
${\cal M}_1, \ {\cal M}_2, \ {\cal M}_3$ are in general complex rank 1 mass matrices.  However only the
difference in their overall phases may be observable. Thus, there are, in principle, two new CP violating phases
in the neutrino sector, in addition to the four phases already fixed by charged fermion masses and mixing angles.
{\em We shall impose the constraint that neutrino Majorana masses $M_i$ are all real.}  This eliminates two
arbitrary phases. We note that the best fits with free phases for $M_i$ are very close numerically to our zero
phase model.

\medskip
\noindent {\it Fitting the low energy data : $\chi^2$ analysis } \medskip

Yukawa matrices in this model are described by seven real parameters \{$\lambda$,  $\epsilon$, $\tilde \epsilon$,
$\sigma$, $\rho$, $\epsilon'$, $\xi $\} and, in general, four phases \{$\Phi_\sigma, \ \Phi_{\tilde \epsilon}, \
\Phi_\rho, \ \Phi_\xi $\}. Therefore, in the charged fermion sector we have 11 parameters to explain 9 masses and
three mixing angles and one CP violating phase in the CKM matrix, leaving us with 2 predictions.\footnote{Of
course, in any supersymmetric theory there is one additional parameter in the fermion mass matrices, i.e.
$\tan\beta$.  Including this parameter, there is one less prediction for fermion masses, but then (once SUSY is
discovered) we have one more prediction.  This is why we have not included it explicitly in the preceding
discussion. } Note, these parameters also determine the neutrino Yukawa matrix. Finally, our minimal ansatz for
the right-handed neutrino mass matrix is given in terms of three additional real parameters\footnote{In
principle, these parameters can be complex. We will nevertheless assume that they are real; hence there are no
additional CP violating phases in the neutrino sector.}, i.e. the three right-handed neutrino masses. At this
point the three light neutrino masses and the neutrino mixing matrix, $U_{PMNS}$, (3+4 observables) are
completely specified. Altogether, the model describes 20 observables in the quark and lepton sectors with 14
parameters, effectively having 6 predictions.\footnote{Note, the two Majorana phases are in principle observable,
for example, in neutrinoless double-beta decay \cite{Petcov:2004wz}, however, the measurement would be very
difficult (perhaps too difficult \cite{Barger:2002vy}).   If observable, this would increase the number of
predictions to 8.}

In addition to the parameters describing the fermion mass matrices, we have to input three parameters specifying
the three gauge couplings: the GUT scale $M_G$ defined as the scale at which $\alpha_1$ and $\alpha_2$ unify, the
gauge coupling at the GUT scale $\alpha_G$, and the correction $\epsilon_3$ to $\alpha_3(M_G)$ necessary to fit
the low energy value of the strong coupling constant. Finally we have to input the complete set of soft SUSY
breaking parameters and the value of $\mu (M_Z)$.\footnote{The reader will find more details on the soft SUSY
breaking parameters in the next section.} All the parameters (except for $\mu (M_Z)$) are run from the GUT scale
to the weak scale ($M_Z$) using two (one) loop RGEs for dimensionless (dimensionful) parameters. At the weak
scale, the SUSY partners are integrated out leaving the two Higgs doublet model as an effective theory. We
require proper electroweak symmetry breaking. Moreover, the full set of  one loop, electroweak and SUSY,
threshold corrections to fermion mass matrices are calculated at $M_Z$. Below $M_Z$ we use 3 loop QCD and 1 loop
QED RG equations to calculate light fermion masses. More details about the analysis can be found in~\cite{bdr} or
\cite{Dermisek:2003vn}.\footnote{The only difference is that in the present analysis we include all three
families of fermions.}  In addition, we self-consistently include the contributions of the right-handed neutrinos
to the RG running between the GUT scale and the mass of the heaviest right-handed neutrino~\cite{neutrino-RGE}.

The $\chi^2$ function is constructed from observables given in Table~\ref{t:fit}. Note that we over constrain the
quark sector. This is due to the fact that quark masses are not known with high accuracy and different
combinations of quark masses usually have independent experimental and theoretical uncertainties. Thus we include
three observables for the charm and bottom quark masses: the $\overline{MS}$ running masses ($m_c(m_c), \
m_b(m_b)$) and the difference in pole masses $M_b - M_c$ obtained from heavy quark effective theory. The same is
true for observables in the CKM matrix.  For example, we include $V_{td}$ and the two CP violating observables
$\epsilon_K$ and the value for $\sin(2\beta)$ measured via the process $B \rightarrow J/\psi \ K_s$. We thus have
16 observables in the quark and charged lepton sectors. We use the central experimental values and one sigma
error bars from the particle data group~\cite{Eidelman:2004wy}. In case the experimental error is less than $0.1
\%$ we use $\sigma = 0.1 \%$ due to the numerical precision of our calculation.

At present only four observables in the neutrino sector are measured. These are the two neutrino mass squared
differences, $\Delta m^2_{31} $ and $\Delta m^2_{21}$, and two mixing angles, $\sin^2 \theta_{12}$ and $\sin^2
\theta_{23}$. For these observables we use the central values and $2 \sigma$ errors from
Ref.~\cite{Maltoni:2004ei}. The other observables: neutrino masses, 1-3 mixing angle and the phase of the lepton
mixing matrix are predictions of the model. We will discuss these later.

\medskip
\noindent {\it Best fit } \medskip

In Table~\ref{t:fit} we present the best fit for quark and lepton masses and mixing for the $\chi^2$ function
constructed from the 20 observables in the quark and lepton sectors and 3 gauge couplings. The input parameters
for this fit are summarized at the top of the table.   One can see that the charged fermion sector is fit very
well with the largest contributions to $\chi^2$ coming from $m_d/m_s$ and $\sin (2 \beta )$.\footnote{The problem
associated with $m_d/m_s$ is directly connected with the Georgi-Jarlskog~\cite{Georgi:1979df} factor of three
explicit in our model, while the problem with $\sin(2 \beta)$ is a result of using factorizable phases in our
Yukawa matrices \cite{Branco:2004ya}.  S.R. thanks G.C. Branco for pointing out the origin of the latter
problem.}

The neutrino observables, on the other hand, are fit very close to their central values, giving a negligible
contribution to $\chi^2$.  Note, our model does not rely on {\em single} right-handed neutrino dominance. In the
case of
SRHND~\cite{Smirnov:1993af,Altarelli:1999dg,Frampton:2002qc,Kuchimanchi:2002yu,King:2002nf,Dorsner:2004qb} one
obtains maximal atmospheric mixing angle as described (see Eqn. \ref{eq:maximal}), but these models (and others)
differ significantly in the way they obtain the large solar mixing angle.  Many models actually abandon the
hierarchical texture of the neutrino Yukawa matrix or they assume a different form for the right-handed neutrino
mass matrix. In ~\cite{Dermisek:2004tx} it was shown that it is {\em impossible} to get the desired ``mild"
hierarchy in $m_{\nu_2}/m_{\nu_3}$ when one of the right-handed neutrinos dominates in strictly hierarchical
models. To fix this, for example in \cite{King:2003rf}, the authors make the (1 2) element comparable to the (2
2) element. But the Yukawa matrix is no longer hierarchical as in Eqn. \ref{eq:ynu}.  It appears that the only
possible way to get the correct neutrino masses, and at the same time have bi-large mixing with Yukawa matrices
of the form in Eqn. \ref{eq:ynu}, is for the contributions of $M_{R_1}$ and $M_{R_2}$ to be comparable (but with
opposite sign). Note that the (2 3) block of the resulting $M_\nu$ does not change when varying the relative
contribution of $M_{R_1}$ and $M_{R_2}$. In ~\cite{Dermisek:2004tx}, a good fit was found with $M_{R_1}$
contributing only -1.5 times more than $M_{R_2}$. In our model, for the best fit, we find that $M_{R_1}$
contributes only -1.2 times more than $M_{R_2}$.\footnote{We define $r_i = ( |Y_n(1,i)|^2 + |Y_n(2,i)|^2 +
|Y_n(3,i)|^2)/M_{R_i}$ as the contribution of $M_{R_i}$ to $M_\nu$. This definition is not unique, there is no
reason to take $|Y_n(1,i)|^2, ...$ because $Y_n(1,i)^2, ...$ is what appears in $M_\nu$. But the latter is
complex, so for a rough order of merit we take the absolute values instead.}  In summary, comparable
contributions of $M_{R_1}$ and $M_{R_2}$ are essential for getting bi-large mixing and an ``observed" neutrino
mass hierarchy (neutrino mass squared differences interpreted as normal mass hierarchy) in hierachical models
with Yukawa matrices of the form Eqn. \ref{eq:ynu}. Finally the contribution of $M_{R_3}$ is, in a model
independent analysis, not very constrained, since we do not know the mass of the lightest neutrino.  However, for
the best fit in our model it turns out that all three right-handed neutrinos contribute comparably, and there is
no single right-handed neutrino dominance. Indeed, forcing the contribution of $M_{R_3}$ to be negligible (by
pushing it all the way to the GUT scale) makes the fit worse, but does not change the fact that we would have
bi-large mixing.

The best fit to fermion masses prefers the region of SUSY parameter space characterized by $\mu, \ M_{1/2} \ \ll
\ m_{16}$ and $- A_0 \ \simeq \  \sqrt{2} m_{10} \ \simeq \ 2 m_{16}$, where $M_{1/2}$ is the universal gaugino
mass, $m_{10}$ is the universal Higgs mass, $m_{16}$ is the universal squark and slepton mass and $A_0$ is the
universal trilinear coupling.\footnote{In addition, we require non-universal Higgs masses at the GUT scale.  For
the best fit we need  $1/2(m_{H_d}^2 - m_{H_u}^2)/m_{10}^2 \sim .07$.  Note this is significantly smaller than
needed in the past~\cite{bdr}.  That is because the RGE running of neutrino Yukawas from $M_G$ to the heaviest
right handed neutrino has been included self-consistently.  As noted in \cite{bdr}, such running was a possible
source for Higgs splitting.  Evidently it can not be the only source.} This is required in order to fit the top,
bottom and tau masses when the third generation Yukawa couplings unify~\cite{bdr}.  Note, third generation Yukawa
unification receives only small corrections from 2-3 mixing in our model. For completeness, predictions for
squark, slepton and Higgs masses for the best fit are presented in Table~\ref{t:susy_and_higgs}. However, it
should be emphasized that the $\chi^2$ function is not very sensitive to changes in the SUSY parameters, as long
as the relations discussed above are approximately satisfied.\footnote{Note, the three input parameters ($\mu, \
M_{1/2}, \ m_{16}$) are not varied when minimizing $\chi^2$.  Moreover, $\chi^2$ is basically flat with increasing
$m_{16}$.} As a consequence of these relations we expect heavy first and second generation squarks and sleptons,
while the third generation scalars are significantly lighter (with the stop generically the lightest). In
addition, charginos and neutralinos are typically the lightest superpartners.  We predict values of $\tan\beta
\sim 50$ and a light Higgs mass of order 120 GeV.    The specific relations between the SUSY breaking parameters
also leads to an interesting prediction for the process $B_s \rightarrow \mu^+ \mu^- $ with branching ratio in
the region currently being explored at the Tevatron.\footnote{This process is sensitive to the CP odd Higgs mass,
$m_A$, which can be adjusted in theories with non-universal Higgs masses.} Furthermore, the neutralino relic
density obtained for our best fit parameters is consistent with WMAP data~\cite{Dermisek:2003vn} and direct
neutralino detection is possible in near future experiments.  Finally, this region maximally suppresses the
dimension five contribution to proton decay~\cite{Dermisek:2000hr} and suppresses SUSY flavor and CP violation in
general. For a detailed analysis of SUSY and Higgs spectra and related phenomenology see Ref.~\cite{bdr} and
\cite{Dermisek:2003vn}.

\medskip
\noindent {\it Additional predictions}
\medskip

Finally let us discuss the predictions in the lepton sector, summarized in Table~\ref{t:neutrino_predictions}. At
this point we would like to emphasize, that non of these observables were included in the $\chi^2$ analysis and
no parameters were constrained or tweaked in order to get the values we present. Therefore these observables are
genuine predictions of the model. However, it should be noted that slightly different values of some of these
observables might be obtained without significant changes in $\chi^2$. We did not study in detail to what extent
each of the observables can be modified without significantly worsening the fit.

The only mixing angle in the lepton sector, which so far has not been measured, is 1-3 mixing. The $2 \sigma$
upper bound from the global fit to neutrino data (with Chooz and solar+KamLAND data contributing comparably to
the fit) is~\cite{Maltoni:2004ei}:
\begin{equation}
\sin^2 \theta_{13} \ < \ 0.031.
\end{equation}
We obtain the best fit value given by \begin{equation} \sin^2 \theta_{13} \simeq 0.0024 \;\; ({\rm or} \; \sin^2
2\theta_{13} \simeq 0.01). \end{equation} This is well below the upper bound and unfortunately too small to be
seen in current or near future experiments. In particular, after three years of running the Double Chooz
experiment will only probe down to $\sin^2 2\theta_{13} \geq 0.03$ \cite{Ardellier:2004ui} and the maximal
sensitivity goal for other future reactor experiments is just at the border of observability with $\sin^2
2\theta_{13} \geq 0.01$~\cite{Goodman:2005ze}. On the other hand, long baseline accelerator experiments, such as
NUMI and T2K, may be able to probe the best fit value~\cite{Mena:2005sa}.

CP violation in the lepton sector given by, for example, the Jarlskog parameter, \bea J = Im (U_{e1} U_{e2}^*
U_{\mu1}^* U_{\mu2} ) = s_{12} c_{12} s_{13} c_{13}^2 s_{23} c_{23} \sin(\delta) , \eea (where we use the
standard parametrization, Eqn.~\ref{eq:PMNS}, with the abbreviated notation $s_{ij} \equiv \sin \theta_{ij}$ and
$c_{ij} \equiv \cos \theta_{ij}$), is proportional to $\sin \theta_{13}$. In spite of the fact that the best fit
suggests a large CP violating phase $\delta \sim \pi/2$, the smallness of $\sin \theta_{13}$ results in $J \sim
0.01$.   CP violation of this magnitude may be observable at long baseline experiments.  For example, the
JPARC-SK experiment has a potential sensitivity to $\sin^2 2\theta_{13} < 1.5\times 10^{-3}$ and $\delta \sim \pm
20^\circ$ and a comparable sensitivity is expected from the ``Off-axis NUMI" proposal~\cite{McKeown:2004yq} .

Majorana neutrino masses are, in principle, observable in processes like neutrinoless double-beta
decay. The effective mass \bea \left< m_{\beta \beta} \right>  = & \left| \ \sum_i  U_{ei}^2 \ m_{\nu_i}  \
\right| = & \left| \ \sum_i \left| U_{ei} \right|^2 \ m_{\nu_i} \ e^{i \alpha'_i} \ \right|  \eea (where
$\alpha'_i = \alpha_i + 2 \delta, \;\; i=1,2$~\cite{Bilenky:2001rz}) is predicted to be of order $2 \times
10^{-4}$ eV which is too low to see in near-future experiments~\cite{Pascoli:2003ke,McKeown:2004yq}. This is a
consequence of the Majorana phases $\alpha_1$ and $\alpha_2$ being almost opposite (see
Table~\ref{t:neutrino_predictions}). The contribution of $m_{\nu_1}$ and $m_{\nu_2}$ is larger than $m_{\nu_3}$
by an order of magnitude and, due to the opposite Majorana phases, these tend to cancel.  This is also evident
from the prediction for the effective electron-neutrino mass observable, relevant
for the analysis of the low energy beta decay of tritium. This mass parameter is unaffected by Majorana phases
and is predicted to be an order of magnitude larger. The observable,
\begin{equation}
m_{\nu_e}^{eff} = \left( \sum_i \left| U_{ei} \right|^2 m_{\nu_i}^2 \right)^{1/2}
\end{equation}
is predicted to be of order $6 \times 10^{-3}$ eV. The current experimental limit is $m_{\nu_e}^{eff} \lesssim
2.5$ eV with the possibility of future experiments, such as KATRIN, reaching bounds on the order of 0.35 eV
\cite{McKeown:2004yq}.  Unfortunately, both mass parameters may be unobservable by presently proposed
experiments.

The best fit values of the heavy right-handed neutrino masses, defined in Eqn.~\ref{eq:rhmass}, are given by \bea
M_{R_1}, \ M_{R_2}, \ M_{R_3} & \approx 10^{10}, \ 10^{12}, \ 7 \times 10^{14} \;\; {\rm GeV} & . \eea  The
lightest Majorana neutrino, $R_1$, is responsible for leptogenesis. The lepton number asymmetry parameter is
given by~\cite{Covi:1996wh,Buchmuller:2005eh}
 \bea \epsilon_1 & =  - \frac{3}{16 \pi} \frac{1}{(Y_\nu^T Y_\nu^*)_{11}} \sum_{i = 2,3} Im\{([Y_\nu^\dagger
Y_\nu]_{1i})^2 \frac{ M_{R_1}}{M_{R_i}}\} &  \eea where $Y_\nu$ is the Dirac neutrino Yukawa matrix\footnote{This
formula is valid for standard model neutrinos.  In the case of non-thermal leptogenesis in supersymmetric
theories, the above value of $\epsilon_1$ is multiplied by an additional factor of 4, taking into account
right-handed neutrino decays to leptons (or sleptons) in the final state with ordinary particles (and
superpartners) in loops.}. This formula is only valid in the limit $M_{R_1} << M_{R_2}, M_{R_3}$. An acceptable
baryon asymmetry, obtained via leptogenesis, requires values of $\epsilon_1 \sim O(10^{-6})$.   In comparison,
for the relevant parameter for non-thermal leptogenesis we find   $4 \times \epsilon_1 \approx - 4.6 \times
10^{-7}$. Note that the baryon asymmetry obtained from leptogenesis comes via electroweak baryon and lepton
number violating interactions preserving  B - L.  Hence the resultant baryon asymmetry $N_B \propto - N_L \propto
-\epsilon_1$. {\em Hence we even obtain the correct sign.}\footnote{In fact, we can obtain either sign. We obtain
the opposite sign with a small increase of $\chi^2$.}

For the case of thermal leptogenesis, the relevant asymmetry parameter is $8 \times \epsilon_1 \approx - .9
\times 10^{-6}$.~\footnote{Since we now also include the asymmetry due to sneutrino decays.}   We note that the
values for the parameters $M_{R_1} \approx 10^{10}$ GeV and $m_{\nu_1} \approx 4 \times 10^{-3}$ eV are in the
regime where the cosmological baryon asymmetry only depends on neutrino
properties~\cite{Buchmuller:2005eh}.\footnote{S.R. would like to thank W. Buchm\"{u}ller for emphasizing this
point.} It does not depend on the initial neutrino abundance nor the initial baryon asymmetry generated by other
mechanisms.  On the other hand, thermal leptogenesis requires the reheat temperature after inflation, $T_R >
M_{R_1} \approx 10^{10}$ GeV. Such a high reheat temperature will generate a gravitino problem, unless the
gravitinos are heavy with mass greater than ${\cal O} (10 $ TeV).

\protect
\begin{table}
\caption[8]{
{\bf The best fit for fermion masses and mixing angles.} \\
   \mbox{Initial parameters: }\ \ \
\ \ (1/$\alpha_G, \, M_G, \, \epsilon_3$) = ($24.98,
\, 2.77 \times 10^{16}$ GeV, $\, -3.2$ \%), \makebox[1.8em]{ }\\
($\lambda, \, \epsilon, \, \sigma, \, \tilde \epsilon, \, \rho, \, \epsilon', \, \xi$) =
($ 0.64, \, 0.046, \, 0.83, \, 0.011, \, -0.053, \, -0.0036, \,
0.12  $),\\
($\Phi_\sigma, \, \Phi_{\tilde \epsilon}, \, \Phi_\rho, \, \Phi_\xi$) =  ($0.618, \, 0.411, \, 0.767, \,
3.673$) rad,
\makebox[6.6em]{ }\\
($m_{16}, \, M_{1/2}, \, A_0, \, \mu(M_Z)$) = ($3500,\, 450, \, -6888.3, \,
247.9$) GeV,\\
($(m_{H_d}/m_{16})^2, \, (m_{H_u}/m_{16})^2, \, \tan\beta$) = ($2.00,  \, 1.71, \, 49.98$) \\
($M_{R_3}, \, M_{R_2}, \, M_{R_1}$) = ($5.8 \times 10^{13}$ GeV, $\, - 9.3 \times 10^{11} $ GeV, $\, 1.1 \times
10^{10} $ GeV) \\


} \label{t:fit}
$$
\begin{array}{|l|c|l|r|}
\hline
{\rm Observable}  &{\rm Data} \ (\sigma) & {\rm Theory} &  {\rm Pull} \\
{\rm (masses \ in \ GeV)}   &  &   &  \\
\hline
\;\;\;G_{\mu} \times 10^5   &  1.16637 \ (0.1 \%) & 1.16635  & < 0.01    \\
\;\;\;\alpha_{EM}^{-1} &  137.036 \ (0.1 \%) & 137.031  &  < 0.01        \\
\;\;\;\alpha_s(M_Z)    &  0.1187 \ (0.002) &   0.1184 &  0.02       \\
\hline
\;\;\;M_t              &  174.3 \ (5.1)   &   175.36  &       0.04 \\
\;\;\;m_b(M_b)          &    4.25 \ (0.25) &  4.252    &  < 0.01               \\
\;\;\;M_b - M_c        &    3.4 \ (0.2) &   3.513  &  0.32              \\
\;\;\;m_c(m_c)  &   1.2 \ (0.2)   &   1.03    &      0.72      \\
\;\;\;m_s              &  0.105 \ (0.025) &    0.114   &   0.13       \\
\;\;\;m_d/m_s          &  0.0521 \ (0.0067) &   0.0627    &   {\bf  2.53}       \\
\;\;\;Q^{-2} \times 10^3  &  1.934 \ (0.334)  &  1.763    &   0.26         \\
\;\;\;M_{\tau}         &  1.777 \ (0.1 \%)  &   1.777   &    < 0.01       \\
\;\;\;M_{\mu}          & 0.10566  \ (0.1 \%) &   0.10566  &  < 0.01      \\
\;\;\;M_e \times 10^3      &  0.511 \ (0.1 \%)&   0.511   &  < 0.01 \\
 \;\;\;V_{us}         &  0.22 \ (0.0026) &   0.2192  &   0.09      \\
\;\;\;V_{cb}         & 0.0413 \ (0.0015) &   0.0407   &    0.15         \\
\;\;\;V_{ub}    & 0.00367 \ (0.00047)  &  0.00332  &   0.56              \\
\;\;\;V_{td} &  0.0082 \ (0.00082) &  0.00819   &     < 0.01   \\
\;\;\;\epsilon_K          &  0.00228 \ (0.000228) &   0.00238  &    0.19       \\
\;\;\;\sin(2\beta) &      0.736 \ (0.049) &  0.6757 & {\bf 1.51}  \\
\hline
\;\;\;\Delta m^2_{31} \times 10^3  &    2.3 \ (0.6) &  2.407 &  0.03  \\
\;\;\;\Delta m^2_{21} \times 10^5   &   7.9 \ (0.6) &  7.866   &  < 0.01  \\
\;\;\;\sin^2 \theta_{12} & 0.295 \ (0.045) &   0.2851  &    0.05  \\
\;\;\;\sin^2 \theta_{23}  &  0.51 \ (0.13) &  0.546  &   0.08  \\
\hline
  \multicolumn{3}{|l}{{\rm TOTAL}\;\;\;\; \chi^2}  & {\bf 6.70}  \\
\hline
\end{array}
$$
\end{table}

\begin{table}
\caption[8]{Predictions for SUSY and Higgs spectra for
the best fit given in Table~\ref{t:fit}.}
\label{t:susy_and_higgs}
$$
\begin{array}{|l|c|}
\hline
{\rm Particle}  & {\rm Mass \ (GeV)} \\
\hline
\;\;\; h              & 121.6  \\
\;\;\; H              & 495.5  \\
\;\;\; A^0            & 499.9  \\
\;\;\; H^+            & 411.5  \\
\;\;\; \chi^0_1       & 174.1  \\
\;\;\; \chi^0_2       & 241.7  \\
\;\;\; \chi^+_1       & 230.7  \\
\;\;\; \tilde g       & 1181.0 \\
\;\;\; \tilde t_1     & 299.4 \\
\;\;\; \tilde b_1     & 748.4 \\
\;\;\; \tilde \tau_1  & 1128.1 \\
\hline
\end{array}
$$
\end{table}

\begin{table}
\caption[8]{Predictions for neutrino masses, $\sin \theta_{13}$ and CP violation in the lepton sector for
the best fit given in Table~\ref{t:fit}.}
\label{t:neutrino_predictions}
$$
\begin{array}{|l|c|}
\hline
\;\;\; m_{\nu_3} \ {\rm (eV)} \;\; &   0.0492    \\
\;\;\; m_{\nu_2} \ {\rm (eV)} \;\; &   0.0096    \\
\;\;\; m_{\nu_1} \ {\rm (eV)} \;\; &   0.0037  \\
\;\;\;\sin^2 \theta_{13}\;\;       &  0.0024        \\
\;\;\; J                                  &     0.0107   \\
\;\;\; \sin \delta      \;\;        &     0.98   \\
\;\;\; \alpha_1 \ {\rm (rad)}      \;\;        & -1.286  \\
\;\;\; \alpha_2  \ {\rm (rad)}     \;\;        & 1.821  \\
\;\;\;  \left< m_{\beta \beta} \right>  \ {\rm(eV)}    \;\;        & 0.00021  \\
\;\;\;   m_{\nu_e}^{eff}   \ {\rm (eV)}    \;\;        & 0.0065  \\
\;\;\; \epsilon_1       \;\;        & -1.16 \times 10^{-7}  \\
\hline
\end{array}
$$
\end{table}

\bigskip

\noindent {\it Conclusions}
\medskip

We construct a simple $SO(10)$ SUSY GUT with $D_3$ family symmetry and an unbroken low energy R parity. The model
describes fermion mass matrices with 14 parameters and gives excellent fits to 20 observable masses and mixing
angles in both quark and lepton sectors, giving 6 predictions. Bi-large neutrino mixing is obtained with
hierarchical quark and lepton Yukawa matrices; thus avoiding the possibility of large lepton flavor violation.
The model naturally predicts small 1-3 neutrino mixing, $\sin \theta_{13} \simeq 0.05$, and a CP violating phase
$\delta$ close to $\pi/2$. Among other interesting predictions is a tiny effective Majorana mass for neutrinoless
double-beta decay. Leptogenesis is also possible with the decay of the lightest right-handed neutrino giving an
acceptable CP violating asymmetry $\epsilon_1$ of order $10^{-6}$, {\em and with the correct sign for the
resultant baryon asymmetry}. We also note that similar models with the non-abelian  symmetry groups $SU(2)$ or
$D_4$, instead of $D_3$, can be constructed~\cite{Dermisek:2005ij}.

The model presented in this letter provides an excellent benchmark for testing supersymmetric GUTs.   At the very
least, this model provides a phenomenological ansatz for fermion Yukawa matrices at the GUT scale, which fits low
energy fermion masses and mixing angles.   There are several possible directions for future research which should
be explored.  First, the model will make predictions for flavor violating processes, such as $\mu \rightarrow e \
\gamma$.  This process, as well as other lepton flavor violating processes, will be tested to much higher
accuracy in the future \cite{meg}.   Secondly, as discussed in the text we obtain acceptable values for the
lepton number asymmetry parameter relevant for leptogenesis, $\epsilon_1$.   Thus it would be interesting to see
if non-thermal leptogenesis via inflaton decay, as discussed by several authors (see \cite{Kyae:2005vg} and
references therein), can be incorporated into this theory.

Finally, can this theory be derived from a more fundamental starting point such as string theory.  Although we
have not shown that this is possible, it is likely that it may be re-written as a five dimensional orbifold GUT
(see \cite{Kim:2004vk}).

\vspace{0.5cm}
{\bf Acknowledgements}

Partial support for S.R. was obtained under DOE grant DOE/ER/01545-863. R.D. was supported, in part, by the U.S.
Department of Energy, Contract DE-FG03-91ER-40674 and the Davis Institute for High Energy Physics.

%

\end{document}